\documentclass[11pt]{article}
\usepackage{moriond,epsfig}
\usepackage{amssymb}

\def\Journal#1#2#3#4{{#1} {\bf #2}, #3 (#4)}

\def\NIMA{{\em Nucl. Instrum. Methods} A}

\def\PLB{{\em Phys. Lett.}  B}

\def\PRD{{\em Phys. Rev.} D}

\def\mco{\multicolumn}

\def\be{\begin{equation}}
\def\ee{\end{equation}}
\def\bea{\begin{eqnarray}}
\def\eea{\end{eqnarray}}

\newcommand{\met}       {\hbox{E\kern-0.5em\lower-0.1ex\hbox{/}}_T}

\begin{document}
\vspace*{4cm}
\title{SEARCH FOR LOW MASS HIGGS BOSON AT THE TEVATRON}

\author{Pierluigi Totaro for the CDF and D0 Collaborations}

\address{Universit\`a di Padova, Dipartimento di Fisica, via Marzolo 8,\\
35131 Padova, Italy}

\maketitle\abstracts{
 We present the current status of searches for a low mass Standard 
Model Higgs boson ($M_H$ below $\lesssim$ 135 GeV/$c^2$)
 using data collected from $p\bar{p}$ collisions at the Fermilab Tevatron
 collider at $\sqrt{s}=$ 1.96~TeV.  A summary of the
latest results from the CDF and D0 collaborations is reported in this paper,
 focusing in particular on ongoing efforts to increase overall search
 sensitivity through improvements to the analysis methods. 
}

\section{Introduction}
The Higgs mechanism is introduced in the Standard Model (SM) to provide mass to fundamental particles, 
through the spontaneous breaking of the electroweak simmetry. This mechanism implies the existence of a 
yet experimentally unobserved scalar particle, the Higgs boson, whose search has represented one of the 
major goals of the high energy physics community over the last decade. The mass of the Higgs boson is a 
free parameter of the theory, but the strong coupling to massive particles allows to constrain its value:
a global fit, which incorporates the measurements of the top quark and W boson masses, as well as additional
 precision electroweak data provided by LEP, SLD and Tevatron experiments~\cite{globalfit}, indicates that a 
light Higgs is preferred, M$_H$=89$^{+35}_{-26}$ GeV/c$^2$, with a 95\% Confidence Level (C.L.) upper limit of
 158 GeV/c$^2$. On the other hand, results from direct searches at LEP~\cite{LEP} set a 95\% lower limit 
of 114.4 GeV/c$^2$.

\begin{figure}
 \begin{minipage}[b]{8cm}
   \centering
   \epsfig{figure=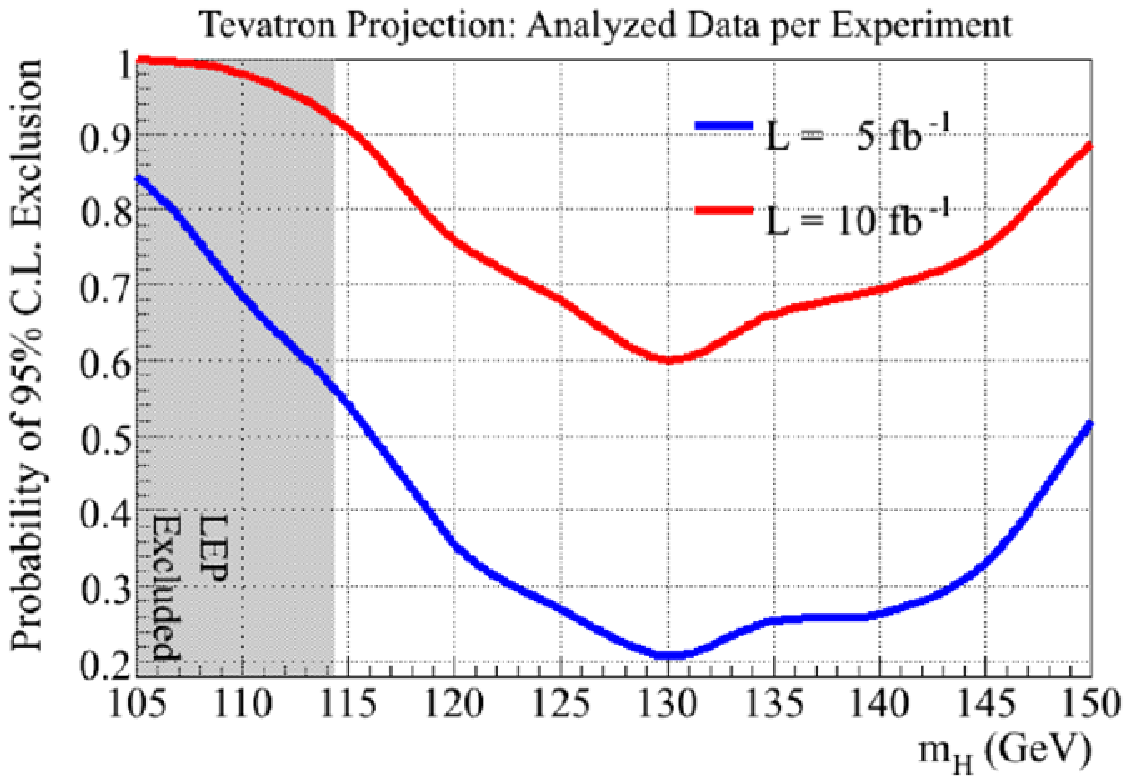,trim=0cm 0cm 0cm 1cm, height=1.8in}
   \caption{Tevatron probability projections of seeing a 2$\sigma$ SM Higgs signal excess, for integrated luminosities of
 5 fb$^{-1}$ and 10 fb$^{-1}$ per experiment, assuming improvements in the analysis techniques.\label{fig:projection}}
 \end{minipage}
  \hspace{2mm}
 \begin{minipage}[b]{8cm}
  \centering
\epsfig{figure=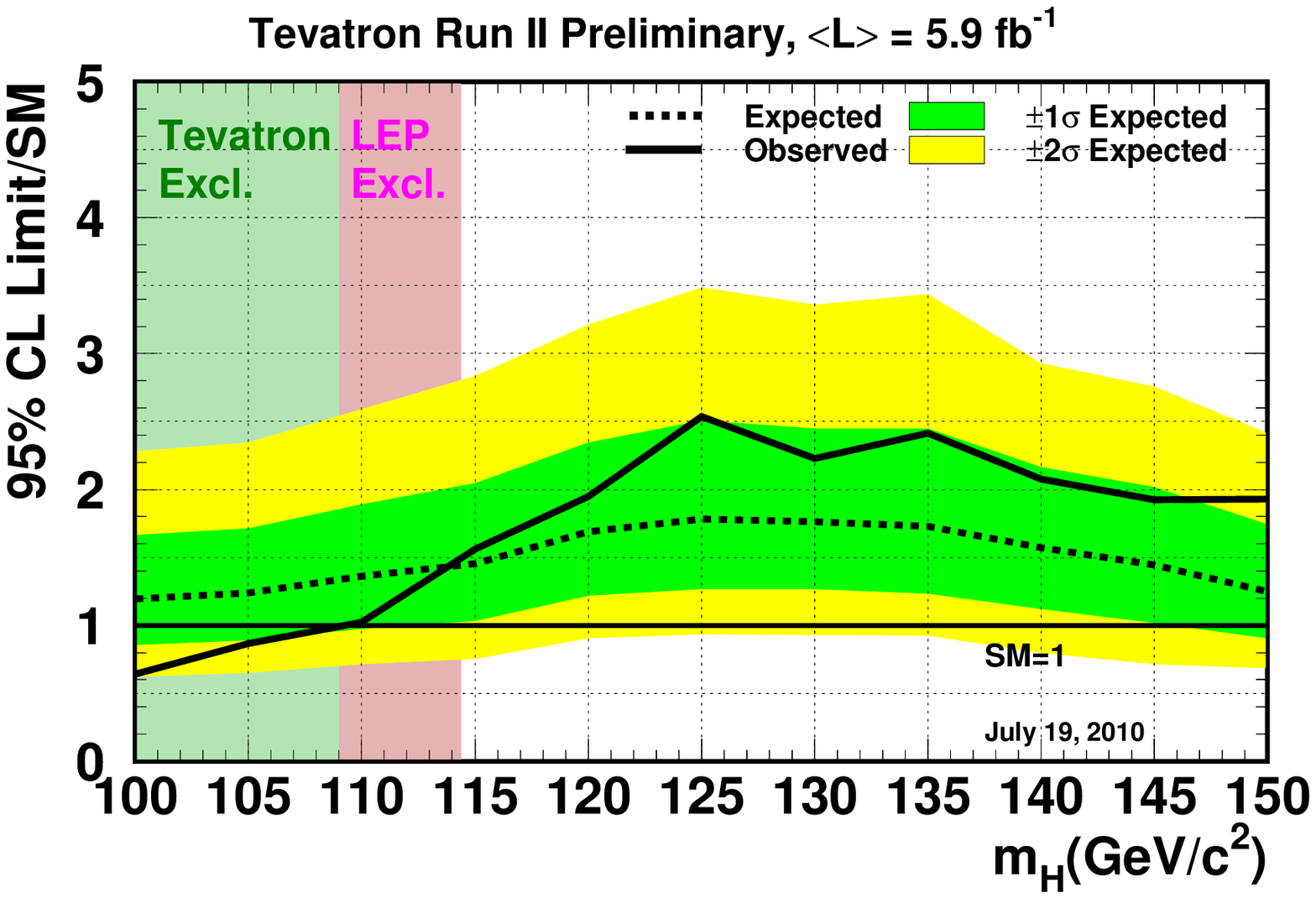,trim=0.5cm 1cm 0cm 0cm, height=2.4in}
   \caption{The observed and expected 95\% C.L. upper limits on the Higgs production cross section, in units of the SM 
theoretical cross section, obtained by combining all CDF and D0 analyses with up to 5.7 fb$^{-1}$ of data.\label{fig:limit}}
 \end{minipage}
\end{figure}

In the last few years CDF and D0 have steadily increased the efforts in extending the potential sensitivity
of their searches: the most recent combined results~\cite{Comb_Winter_2011} exclude the existence of the Higgs boson 
with a mass between 158 and 173 GeV/c$^2$. This interval is expected to further extend, as well
as new data will be included. However, a substantial chance to make a signal observation
or set an exclusion in the entire explored mass range (100$\div$200 GeV/c$^{2}$) will require several improvements 
in the analysis methods, beyond the increase of statistics provided by the end of Tevatron operations: a projection 
of the probability of seeing a 2$\sigma$ excess, for an integrated luminosity of 10 fb$^{-1}$ per experiment, 
calculated assuming 30$\div$40\% of sensitivity increase in the analysis techniques with respect
to Summer 2010 results, is reported in figure~\ref{fig:projection}. 

As of this paper, the observed upper limit at the reference 
mass of 115 GeV/c$^2$ is 1.58 times the predicted SM cross section (figure~\ref{fig:limit}): this value refers to the CDF
and D0's combined measurements with up to 5.7 fb$^{-1}$ of data~\cite{Comb_Summer_2010}. The plan of the two collaborations 
is to come out with a new more stringent combined limit in the low mass region by Summer 2011, when several search 
channels will almost double the analyzed integrated luminosity. A summary of the latest public results in the low mass
 Higgs boson searches is given in this paper, focusing on the most significant improvements which are being implemented 
and will allow to reach the best sensitivity in the next Tevatron combination.

\section{Experimental apparatus}

A detailed description of the Tevatron collider and CDF and D0 detectors can be found elsewhere~\cite{CDF,D0}.
The accelerator provides $p\bar{p}$ collisions at $\sqrt{s}$ = 1.96 TeV with
stable and well performing operating conditions: as of May 2011 about 60 pb$^{-1}$ are  
produced per week, with a typical instantaneous luminosity of 3$\times$10$^{32}$ cm$^{-2}$s$^{-1}$; 
since the beginning or Run II, over 10 fb$^{-1}$ of data have been delivered at the two 
collision points, and more than 8 fb$^{-1}$ were recorded and made available for the
 analyses by each experiment. Tevatron collisions are scheduled to stop in September 2011 and we expect that an additional 
2 fb$^{-1}$ of data will be delivered by that date.

\section{Low Mass Higgs Boson at the Tevatron}
At the Tevatron center of mass energy, the dominant Higgs production mode is represented by gluon-gluon fusion,
gg$\rightarrow$H, followed by the associated production with a W or a Z boson, $q\bar{q}\rightarrow$(W/Z)H, and
the vector boson fusion, $qq\rightarrow q$H$q$. Depending on the mass,
 the inclusive predicted cross section in the 100$\div$200 GeV/c$^{2}$ interval ranges from about 2 to 0.7 pb: the achievable 
signal yield is therefore particularly small if compared to the main SM background processes, which are several orders of 
magnitude larger.

The Higgs search is particularly challenging for M$_H \lesssim$ 135 GeV/c$^2$, where the decay mode into b quarks 
becomes dominant ($\sim$73\% at M$_H$=115 GeV/c$^2$), making difficult the investigation of the direct production:
 although being the most abundant, the gg$\rightarrow$H$\rightarrow b\bar{b}$ process is indeed experimentally 
prohibitive because of the overwhelming non-resonant multijet background. It is then preferred to consider the 
associated production, whose cross section is smaller of one order of magnitude, but where the leptonic decays 
of the W and Z boson provide cleaner signatures, easy to trigger on and with a great reduction of the QCD background. 
The most sensitive channels are represented by WH$\rightarrow l\nu b\bar{b}$, ZH$\rightarrow\nu\bar{\nu}b\bar{b}$ and 
ZH$\rightarrow llb\bar{b}$, where the Higgs boson is detected through the reconstruction of the jets originating from 
the b quark hadronization.

Many other additional channels, although less powerful, are considered since they provide a sizeable contribution to the overall
sensitivity: these include the all-hadronic associated production, where the W and Z bosons are searched 
in their hadronic decay, and the low branching ratio (B.R.) decay into a pair of tau leptons or a pair of photons. 

No single channel provides by itself the sensitivity to discover the Higgs boson. The best strategy is to perform 
dedicated analyses exploiting the specific topological features of the different final states and then combine the 
results into one single measurement. In order to maximize the sensitivity and optimize the analysis techniques, 
each channel can be further split into subcategories according to the lepton types or the jet multiplicity in the event 
selection.

\section{Analysis strategies}
\subsection{Acceptance optimization}
One of the main challenges in the Higgs searches is represented by the need to increase as much as possible the total 
signal acceptance: Tevatron experiments are pursuing this target by including new triggers in the online selection, 
by relaxing the kinematic cuts and by implementing additional lepton categories or more sophisticated identification 
algorithms in the event reconstruction. The larger explored phase space requires nevertheless an accurate understanding 
of the selected data sample, whose composition has to be well described by the background modelings.

\begin{figure}[h!]
\begin{center}
\epsfig{figure=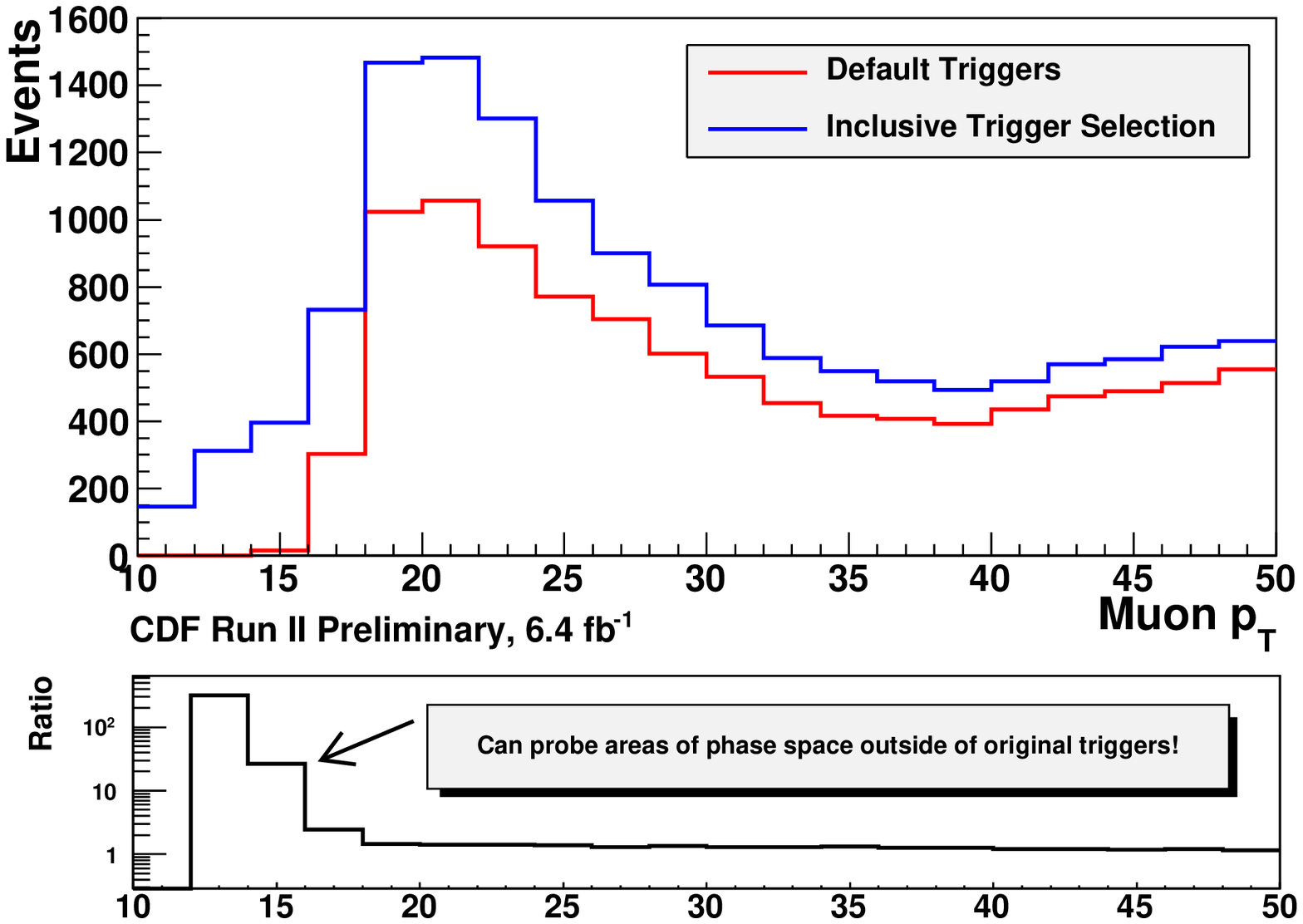,height=1.8in}%
\epsfig{figure=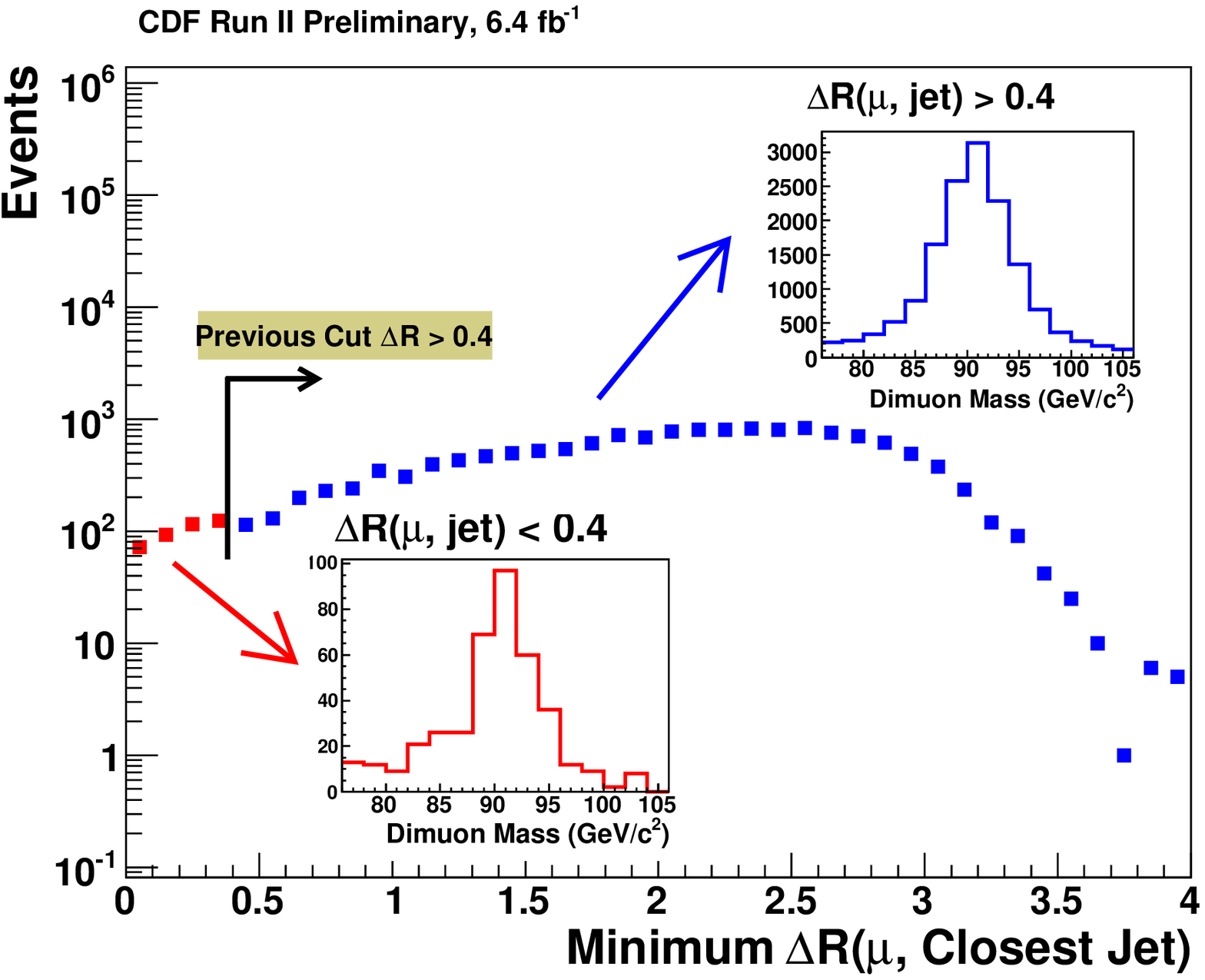,height=1.9in}
\caption{Acceptance increase in the CDF's ZH$\rightarrow \mu \bar{\mu} b \bar{b}$ search. Left: implementation
of an inclusive trigger selection. Since no specific cuts on muon candidates are applied, a larger fraction of events
is recored, compared to the standard high p$_T$ ($\ge$18 GeV/c) muon trigger. Right: removal of the spatial separation cut
($\Delta$R=$\sqrt{{\Delta\eta}^2+{\Delta\varphi}^2}>$0.4) between the muon and the closest jet.
\label{fig:CDF_ZHmumubb}}
\end{center}
\end{figure}

An example of the potential gain provided by the increased event acceptance is given by the ongoing update of CDF's 
ZH$\rightarrow \mu\mu b \bar{b}$ search~\cite{ZHllbb}: the preliminary results, obtained by employing a novel muon identification
based on a neural network (NN) algorithm, as well as an extended kinematic selection, as described in figure~\ref{fig:CDF_ZHmumubb},
 indicate a sensitivity improvement of the order of 30$\div$60\% beyond the luminosity scaling, in the 100$\div$150 GeV/c$^2$
mass range.

\subsection{b-quark identification}
When considering final states including b quarks, one fundamental ingredient is the capability of distinguishing 
jets originated from b quarks from those coming from gluons, light or c quarks. 

Both CDF and D0  have developed specific ''b-tagging'' algorithms, which exploit the relatively long lifetime
 of b-hadrons and the high position resolution of the silicon detectors. Different approaches are followed: 
CDF's SecVtx~\cite{SecVtx} is based on the reconstruction of the b-hadron secondary vertex, obtained by fitting 
the tracks displaced from the interaction point; CDF's JetProb~\cite{JetProb} uses the distribution of the track 
impact parameters, with respect to the primary vertex, to build a probability that a jet contains a b-hadron;
More sophisticated algorithms adopted by both CDF and D0 are based on NNs~\cite{btag_NNCDF,btag_NND0} and Boosted 
Decision Trees (BDT)~\cite{ZHnnbb} and combine the information provided by different taggers, with the discriminating
 power of additional variables, including those related to the leptonic decay of b-hadrons inside the jet. 
These multivariate methods benefit of the correlations among the input variables, which help in increasing the 
signal to background separation; in addition, they have the advantage to provide continuos outputs instead of a 
simple binary one. This allows to easily modify the definition of a b-tagged jet, by changing the cut on the
output distributions, and then alternatively maximize the sample purity or increase the signal acceptance of the
 analysis selection. 
Typical b-tagging efficiencies are 40$\div$70\%, with a corresponding light flavour jet mistag rate of 0.5$\div$3\%.

\subsection{Multivariate techniques}

Given the small signal to background ratio, the analyses employ multivariate techniques in order to exploit all the 
event information, by collecting multiple distributions into a single and more powerful discriminating variable: 
 the preferred methods are based on NNs, BDTs and matrix elements (ME). The search sensitivity usually increases 
by about 20\% with respect to simply using one single kinematic distribution as discriminator. 

The reliability of these techniques depends on the goodness of the background modeling for the input variables, which
 need to be carefully verified in dedicated control samples.

\section{Results}

In table~\ref{tab:exp} we summarize the expected and observed 95\% C.L. upper limits for the different CDF and D0 search 
channels. More information can be found in the references and in the web pages of the two experiments~\cite{www_cdf,www_D0}. 
The items marked with an asterisk refer to the analyses which were updated since Summer 2010 and for which a more detailed 
description is given here. 

\begin{table}[t]                                                                 
\caption{Observed and expected upper limits at 95\% C.L. on the Higgs boson production cross section, at the reference mass
 of 115 GeV/c$^2$, for the CDF and D0 experiments as of May 2011. 
\label{tab:exp}}                         
\vspace{0.4cm}                                                                   
\begin{center}                                                                   
\begin{tabular}{|l|c|c|c|c|c|c|}                                                       
\hline
 & \mco{3}{|c|}{CDF} & \mco{3}{|c|}{D0}\\
\cline{2-7}
Channel  &  $\mathcal{L}$  & Exp.limit &  Obs.limit &  $\mathcal{L}$  & Exp.limit &  Obs.limit\\                          
& [fb$^{-1}$]& [$\sigma$/$\sigma$(SM)] &[$\sigma$/$\sigma$(SM)] &[fb$^{-1}$] &[$\sigma$/$\sigma$(SM)] & [$\sigma$/$\sigma$(SM)]\\ 
\hline                                                                 
WH$\rightarrow l\nu b\bar{b}$~\cite{WH} &5.7 & 3.5& 3.6&5.3 & 4.8&4.1\\
ZH$\rightarrow llb\bar{b}$~\cite{ZHllbb} &5.7 & 5.5& 6.0& 6.2& 5.7& 8.0\\
ZH$\rightarrow \nu\nu b\bar{b}$~\cite{ZHnnbb} &5.7 & 4.0& 2.3& 6.2*& 4.0&3.4\\
VH/VBF$\rightarrow$ $b\bar{b}$+jets~\cite{CDF_allhadronic} &4.0 & 17.8& 9.1& -& -&-\\
H$\rightarrow$ $\tau\tau$+jets~\cite{tautau} &6.0* & 15.2& 14.7& 4.3*& 12.8&32.8\\
H$\rightarrow \gamma\gamma$~\cite{gg} & 4.2& 20.8& 24.6& 8.2*& 11.0&19.9\\
\hline
\end{tabular}                                                                    
\end{center}                                                                     
\end{table}                                                                      

\subsection{ZH$\rightarrow \nu\nu b\bar{b}$}
The signature of this search is based on two b-jets plus an unbalance of transverse energy ($\met$) due to the undetected 
neutrinos, coming from the Z boson invisible decay. The analysis is also sensitive to the WH$\rightarrow l\nu b\bar{b}$ channel, 
when the charged lepton from the W escapes the detection. CDF and D0 apply similar event selections and search strategies:
 they both require large $\met$ and 2 or 3 jets, at least one of them b-tagged. NNs (CDF) and BDTs (D0) are implemented to 
reduce the main background process, represented by QCD multijet production, with $\met$ coming from jet energy mismeasurements. 
A second discriminant is then used to separate the signal from the remaining sources of background. 

\begin{figure}[h!]
\begin{center}
\epsfig{figure=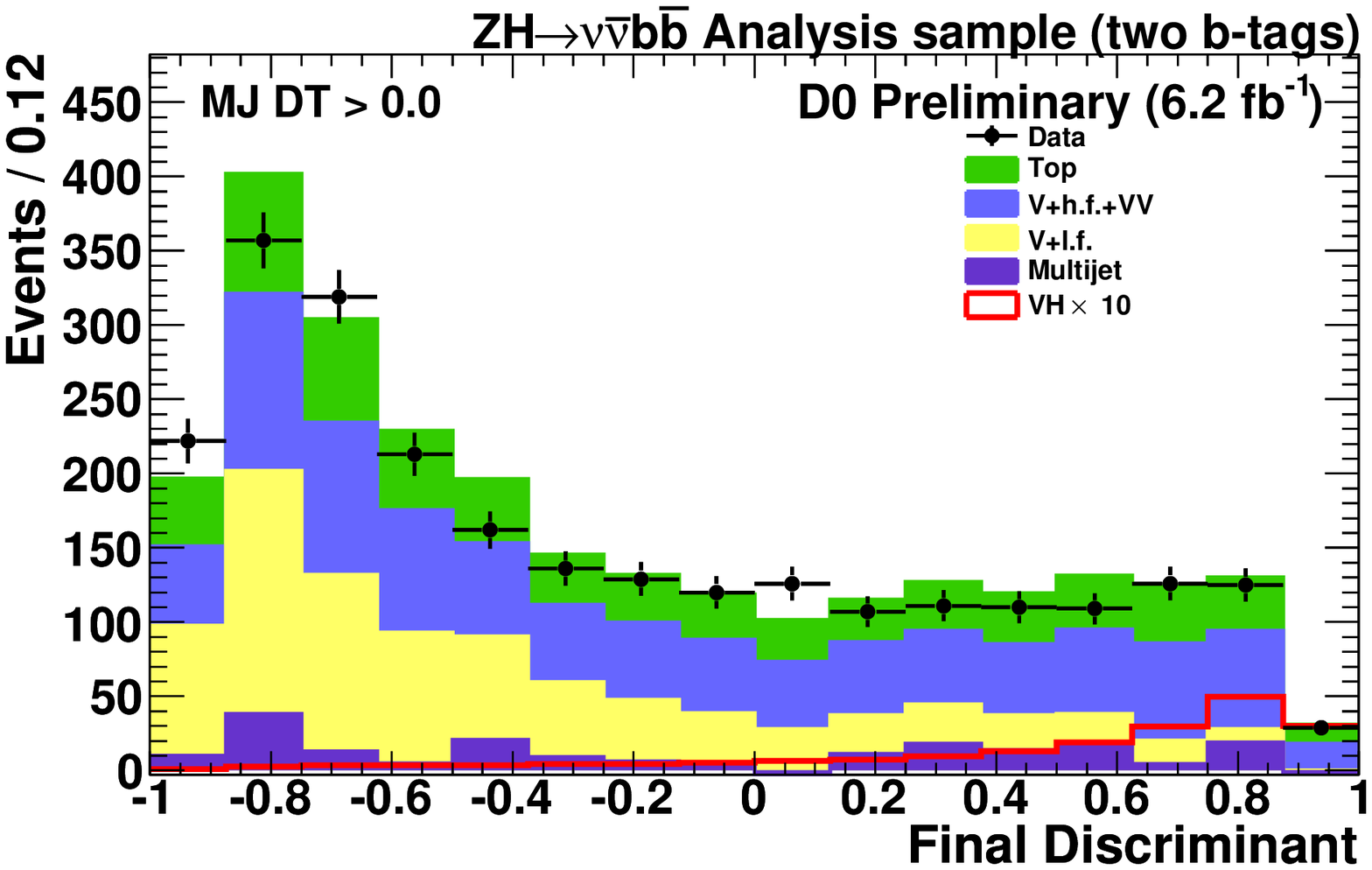,height=1.9in}%
\epsfig{figure=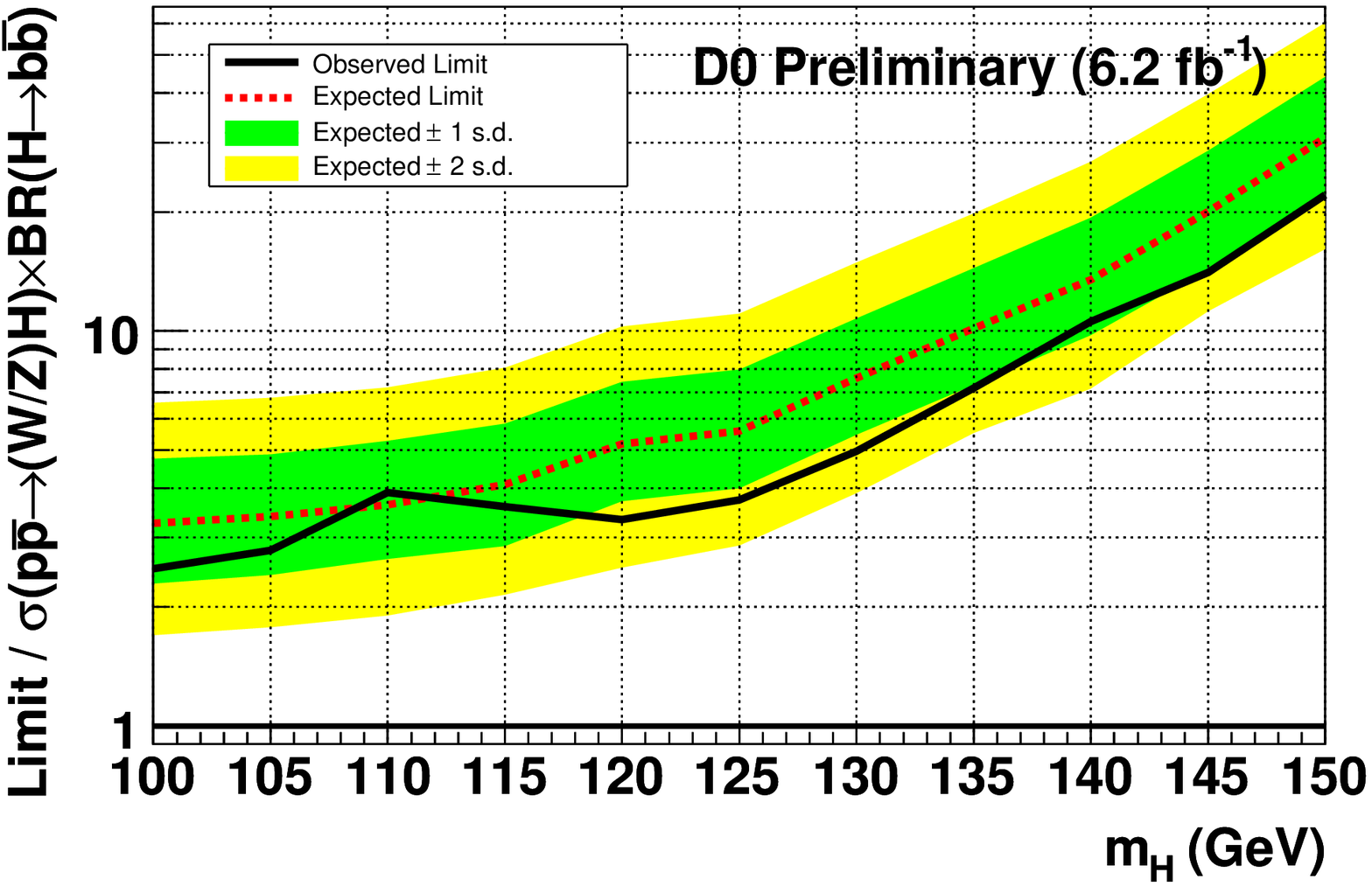,height=1.9in}
\caption{D0's ZH$\rightarrow \nu\bar{\nu} b\bar{b}$ search. Left: final discriminant distribution for the
double b-tag channel, in the Higgs mass hypothesis of 115 GeV/c$^2$. Right: observed and expected upper limits on 
the Higgs boson production cross sections, as a function of the Higgs mass.\label{fig:D0_ZHnunubb}}
\end{center}
\end{figure}

The D0 latest search update has significantly increased the sensitivity thanks to the acceptance gain provided by loosening 
the b quark identification requirements, followed by a more clever use of the b-tagger output information. The latter has been 
employed as additional input variable for the final multivariate algorithm, thus improving the separation between signal and
background. This new approach results in a 14\% improvement in the expected limit compared to the previous version of the analysis.
The final distribution for events containing two b-tagged jets, and the corresponding observed and expected upper limits 
on the Higgs boson production cross section, as a function of the mass, are shown in figure~\ref{fig:D0_ZHnunubb}.

\subsection{H$\rightarrow \tau\tau$+jets}

The B.R. of H$\rightarrow \tau\tau$ is one order of magnitude smaller than H$\rightarrow b\bar{b}$, but the contribution 
of this search is significant, since several production modes can be simultaneously investigated. In particular, the gluon 
fusion becomes accessible thanks to the selection of the leptonic decay of one of the two taus, which considerably reduces 
the multijet background. The requirement of jets in the final state further increases the signal to background ratio and
 optimizes the search for the vector boson fusion process and the associated production, where the W and Z are allowed to 
decay hadronically. However, the significance of this channel is affected by the similarity of the H$\rightarrow \tau\tau$ 
signal with the irreducible Z$\rightarrow \tau\tau$ background, both characterized by a resonant tau pair in the final state. 
One additional challenge is represented by the hard discrimination of real hadronically decaying taus from quark/gluon jets:
 CDF and D0 employ identification algorithms based on BDTs and NNs, respectively.

\begin{figure}
 \begin{minipage}[b]{8cm}
   \centering
   \epsfig{figure=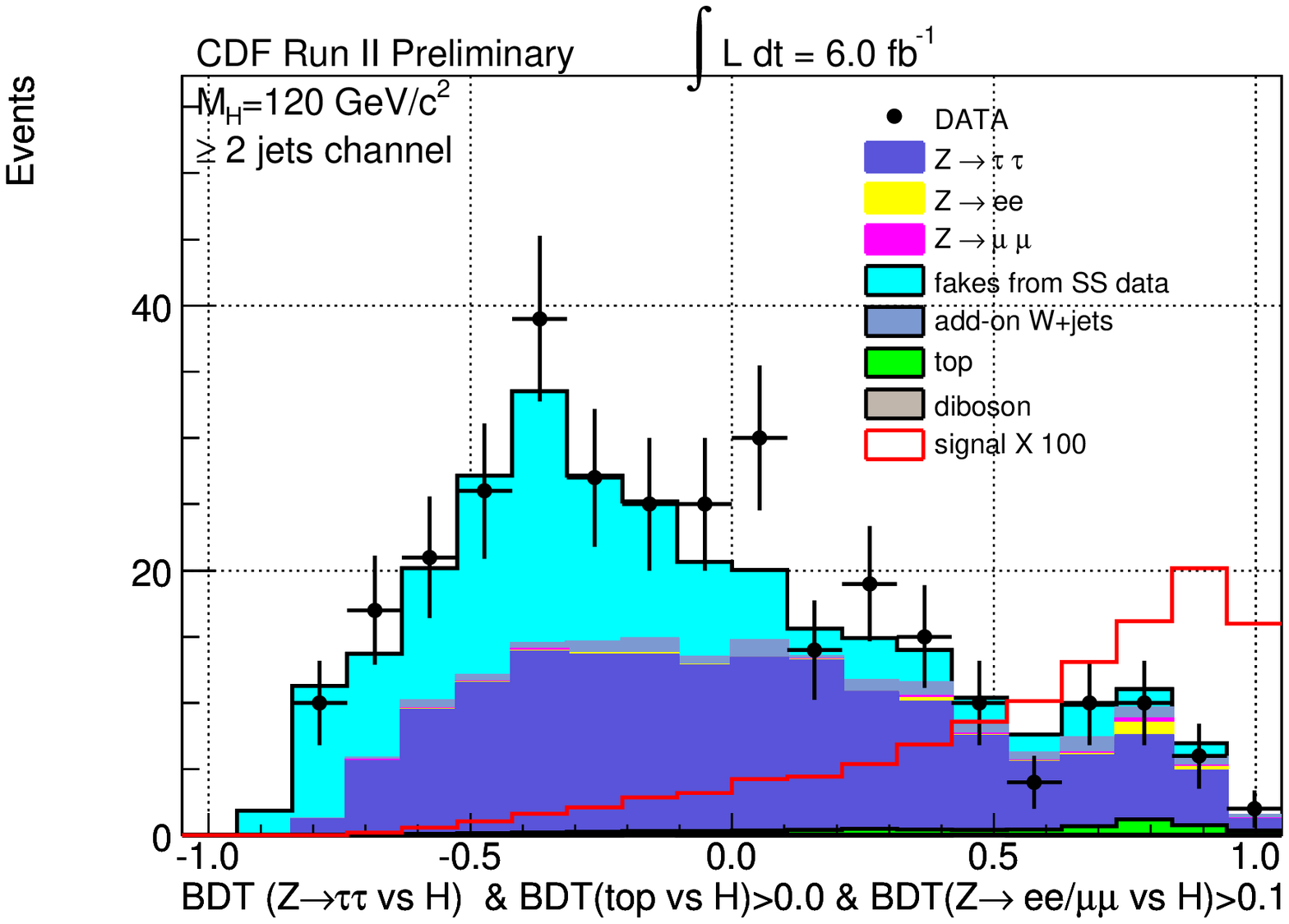,height=2in}%
   \caption{\label{fig:CDF_tautau}CDF's H$\rightarrow \tau\tau$+jets search: final discriminant distribution in the Higgs mass 
hypothesis of 120 GeV/c$^2$.}
 \end{minipage}
  \hspace{2mm}
 \begin{minipage}[b]{8cm}
  \centering
  \epsfig{figure=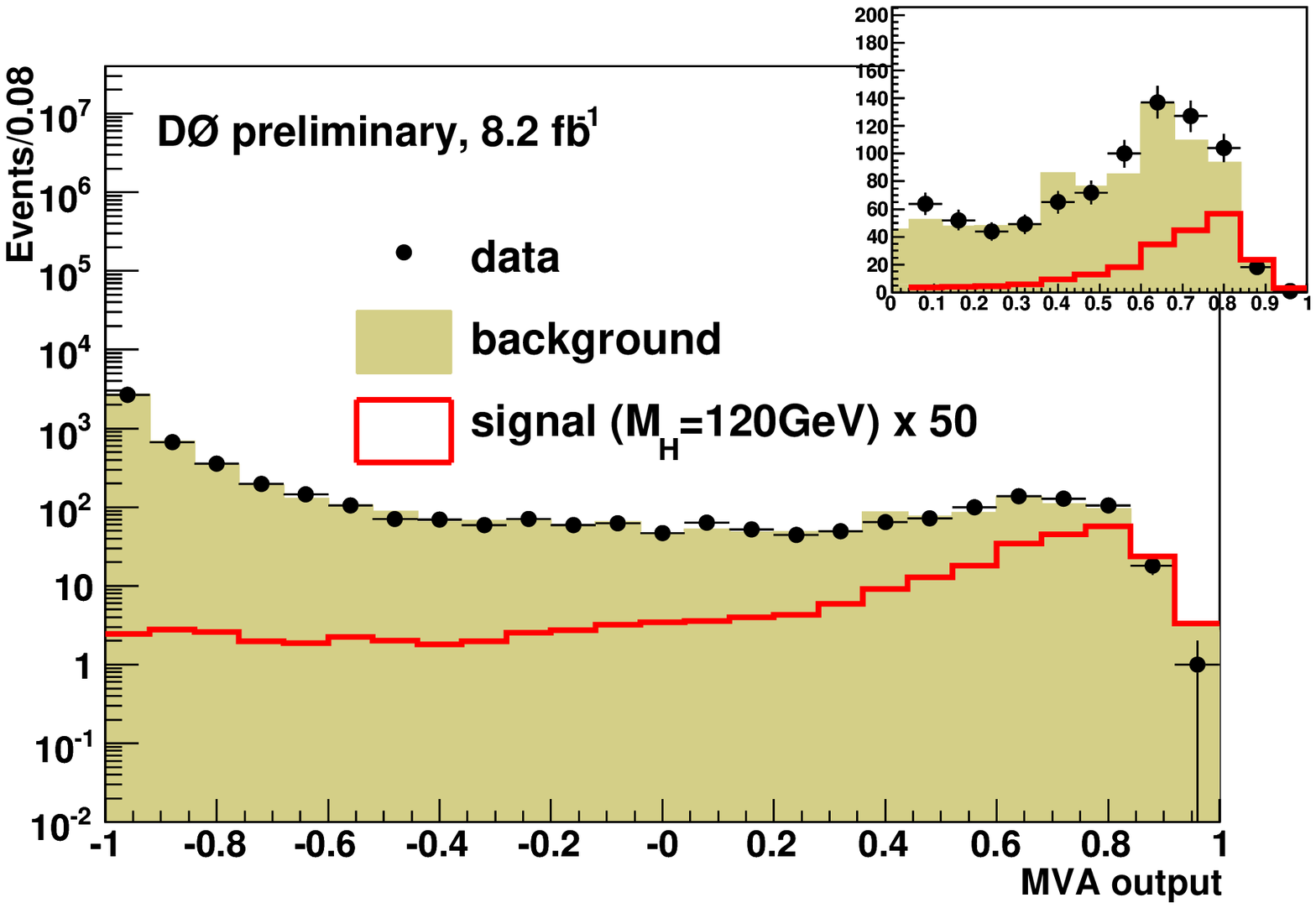,height=2in}
   \caption{\label{fig:D0_Hg}D0's H$\rightarrow \gamma\gamma$ search: final discriminant distribution in the Higgs mass hypothesis
 of 120 GeV/c$^2$.}
 \end{minipage}
\end{figure}

Both the experiments have recently presented an update of their searches, where the most relevant improvements are related 
to the refined multivariate techniques adopted to build the final discriminant. The best separation between signal and 
background is achieved by following a two stage procedure: first several independent BDTs are trained to distinguish the Higgs 
from the principal sources of background; the different outputs are then combined into one single distribution, chosen to 
maximize the sensitivity of the search. Figure~\ref{fig:CDF_tautau} shows the CDF final discriminant for events containing 2 
or more jets in the final state.


\subsection{H$\rightarrow \gamma\gamma$+X}
The diphoton final state suffers from a very low B.R., but it is interesting because the photon
identification efficiency and the energy resolution are much better then that of b-jets, and the narrow M$_{\gamma\gamma}$ mass peak
 can be exploited to reduce backgrounds. The selection is based on the requirement of two high E$_T$ central photons. 
The dominant background is the direct SM diphoton production, followed by events with misidentified electrons and jets. 
CDF sets a limit by looking for a peak resonance in the M$_{\gamma\gamma}$ distribution; D0 has recently implemented a BDT which 
collects five kinematic variables, bringing an improvement of the sensitivity of about 20\% with respect to the luminosity increase 
from the previous stage of the analysis.

\section{Conclusions}
We presented the latest results on the Tevatron searches for a low mass SM Higgs boson. 
The update of the CDF and D0's combination, currently in progress, will benefit from the ongoing 
efforts described in this paper to increase the performances beyond the luminosity scale: the projections shown in 
figure~\ref{fig:projection} suggest that, with the full data expected by the end of Run II, accompained by the suitable 
improvements in the analysis techniques, the Tevatron could reach the sensitivity to exclude the presence of the SM Higgs
 in the entire explored mass range below 150 GeV/c$^2$, with a sizeable chance to set a 3$\sigma$ evidence of its existence.

\end{document}